\documentclass[a4paper]{jpconf}
\usepackage{graphicx}
\begin{document}
\title{Solving the $R_{AA}\otimes v_2$ puzzle}

\author{Jacquelyn Noronha-Hostler}

\address{Department of Physics, University of Houston, Houston TX 77204, USA}

\ead{jakinoronhahostler@gmail.com}

\begin{abstract}
For the past ten years $R_{AA}(p_T)$, the nuclear modification factor that encodes the suppression of high $p_T$ particles due to energy loss within the medium was fairly well described by many theoretical models. However, the same models systematically underpredicted the high $p_T$ elliptic flow, $v_2$, which is experimentally measured as the correlation between soft and hard hadrons. All previous calculations neglected the effect of event-by-event fluctuations of an expanding viscous hydrodynamical background as well as the soft-hard flow harmonic correlations in the experimentally measured $v_2$.  In this talk I show how event-by-event viscous hydrodynamics (computed using the v-USPhydro code) coupled to an energy loss model (BBMG) is able to simultaneously describe soft physics observables as well as the high-$p_T$ $R_{AA}$ and $v_2$.  Suggestions for future more differential calculations at the LHC run2 are made to explore soft-hard flow correlations. 
\end{abstract}

\section{Introduction}

Two key signatures of the Quark Gluon Plasma- collective flow and jet quenching- have been thoroughly measured experimentally over the years at the Relativistic Heavy Ion Collider RHIC at BNL and the Large Hadron Collider LHC at CERN.  Initially, it was thought that the soft physics regime (low transverse momentum $p_T$ where collective flow is measured) and the hard physics regime (high $p_T$ where jet physics is relevant) could be tackled separately.  The early measurements of high $p_T$ elliptic flow \cite{Adare:2010sp} demonstrated flow was possible even within the hard sector though theoretical calculations \cite{Wang:2000fq,Gyulassy:2000gk,Shuryak:2001me} taking into account only energy loss physics and simplified assumptions for the medium evolution were not able to reproduce the large measured $v_2$ and, in fact, the difficulties in simultaneously describing $R_{AA}\otimes v_2$ remained for years to come (see the discussion and references in \cite{Betz:2014cza,Xu:2014tda}).  

\begin{figure}[h]\centering
\includegraphics[width=26pc]{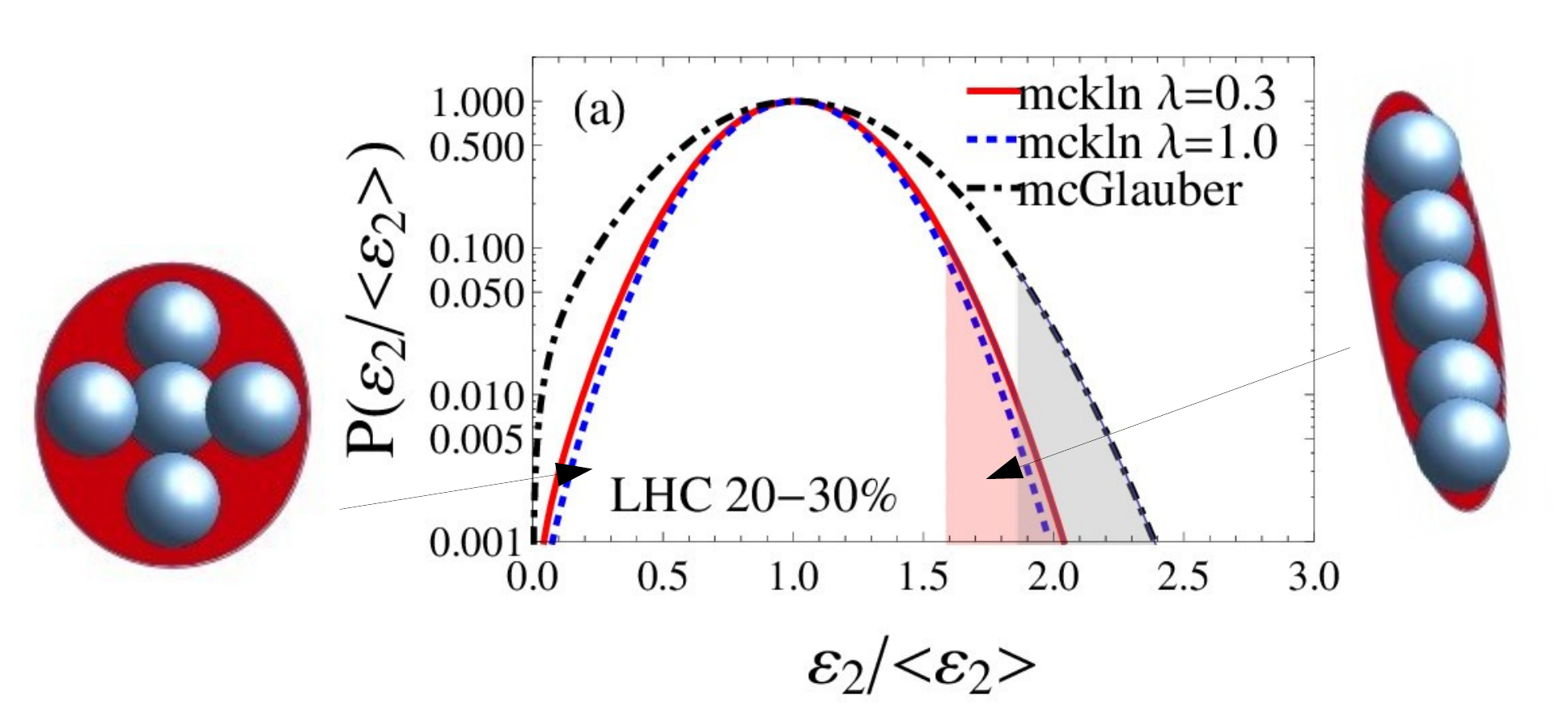}\hspace{2pc}%
\caption{\label{flowdis}Distributions of $\varepsilon_2$ at LHC for MCGlauber and MCKLN (with two different smoothing scales) produce initial conditions that are at one end very circular and at the other end elliptical.}
\end{figure}
 
Meanwhile, comparisons of relativistic hydrodynamical models to collective flow measurements had a breakthrough circa 2010 \cite{Alver:2010gr} with the inclusion of event-by-event calculations \cite{Paiva:1996nv,Gyulassy:1996br,Aguiar:2001ac,Socolowski:2004hw,Andrade:2006yh} to make comparisons to nonzero triangular flow measurements \cite{Alver:2010gr}. Event-by-event fluctuating initial conditions imply that within the same centrality class (i.e. holding the density constant) there is a wide range of initial shapes (eccentricities $\varepsilon_{n}$) that  relativistic hydrodynamics map into the final flow harmonics \cite{Gardim:2011xv,Gardim:2014tya}.  In Fig. \ref{flowdis} the distribution of $\varepsilon_2$ are shown for MCGlauber and MCKLN \cite{Drescher:2006ca} (smoothing out initial energy density fluctuations to $\lambda=0.3$ fm and $\lambda=1$ fm- see \cite{Noronha-Hostler:2015coa} for more details).  Regardless of the initial conditions, a wide range of initial elliptical shapes are produced (and more complex shapes such as ``triangles", ``squares" etc are also produced), which experimentally are proven via $v_n$ distributions \cite{Aad:2013xma}.  

\begin{figure}[h]\centering
\includegraphics[width=20pc]{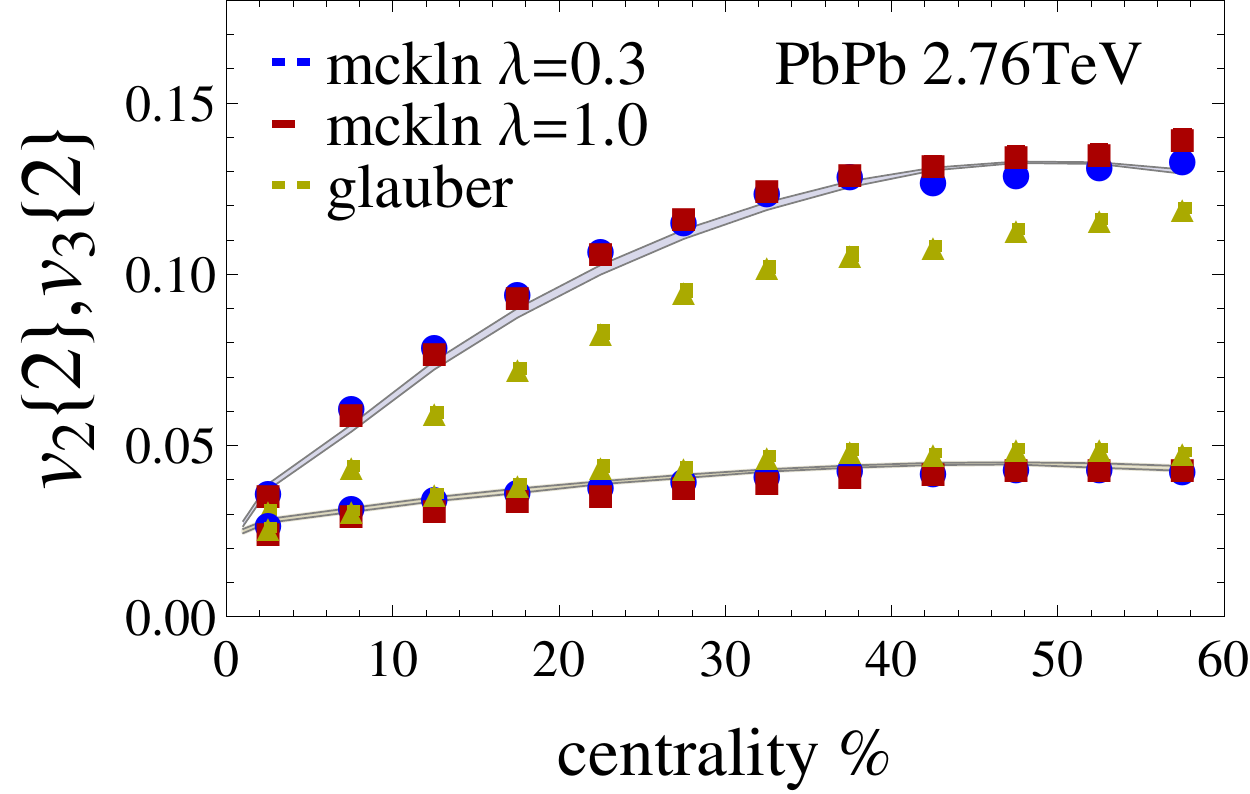}\hspace{2pc}%
\caption{\label{softv2} $v_2\{2\}$ and $v_3\{2\}$ for MCGlauber and MCKLN (with two different smoothing scales) at LHC.}
\end{figure}
Event-by-event viscous hydrodynamical models have been enormously successful at describing soft physics observables such as the cumulants of flow harmonics (as shown in Fig.\ \ref{softv2}) as well as even making predictions for the highest LHC energies \cite{Noronha-Hostler:2015uye}.  While there has been some progress made in combining the knowledge of hydrodynamical backgrounds \cite{Betz:2015mlf,Crkovska:2016flo} and fluctuating initial conditions \cite{Zhang:2012ha} with energy loss models, it was not until \cite{Noronha-Hostler:2016eow} that full event-by-event viscous hydrodynamical backgrounds (v-USPhydro  \cite{Noronha-Hostler:2013gga,Noronha-Hostler:2014dqa}) were combined with an energy loss model (BBMG  \cite{Betz:2014cza,Betz:2011tu,Betz:2012qq}).  The effects of event-by-event fluctuations as well as the correlation between soft and hard flow harmonics in \cite{Noronha-Hostler:2016eow} provided the key step needed in order to solve the longstanding $R_{AA}\otimes v_2$ puzzle.  

\section{Event-by-event viscous hydrodynamics+energy loss}

In \cite{Noronha-Hostler:2016eow} there were two crucial differences between all previous calculations of $v_2(p_T>10 GeV)$:
\begin{enumerate}
\item Full event-by-event viscous hydrodynamical backgrounds were fed into the energy loss model to obtain $v_2^{hard}$.
\item $v_2^{exp}(p_T)$ from Eq. (\ref{eqn:e2exp}) was calculated correlating the soft integrated $v_2^{soft}$ and hard $v_2^{hard}(p_T)$ to simulate the two particle (one soft, one hard) correlations used to measure elliptical flow experimentally.
\end{enumerate}
The soft $v_2^{soft}$ was taken using the best fitting parameters in the soft sector, see \cite{Noronha-Hostler:2016eow} for more details.  The $v_2^{hard}$ (the 2nd Fourier coefficient of $R_{AA}(p_T,\phi)$) is computed using the BBMG code \cite{Betz:2014cza,Betz:2011tu,Betz:2012qq} in the pQCD-like scenario.  In order to make comparisons to experimental data then the soft and hard flow harmonics are correlated via
\begin{equation}\label{eqn:e2exp}
v^{exp}_n(p_T)=\frac{\langle v^{soft}_n\,v_n^{hard}(p_T)\cos\left[n\left(\psi^{soft}_n-\psi^{hard}_n(p_T)\right]\right) \rangle}{\sqrt{\left\langle \left(v^{soft}_n\right)^{2}\right\rangle}},
\label{softhard}
\end{equation}
where $\langle\ldots\rangle$ denote event averages and $\psi^{soft}_n$ and $\psi^{hard}_n(p_T)$ are the soft event plane angle and hard event plane angle, respectively, as was discussed in \cite{Luzum:2013yya}.  Note that if one assumes a smoothed, averaged hydrodynamical background then there is only one ``event" and Eq.\ (\ref{eqn:e2exp}) reduces down to $v^{exp}_n(p_T)|_{(ic=avg)}=v_n^{hard}(p_T)$.

The enhancement of $v_2$ due to the soft-hard $v_n$ correlation in Eq.\ (\ref{eqn:e2exp}) becomes clear if one considers linear response (a good approximation as long as one is not considering peripheral collisions \cite{Noronha-Hostler:2015dbi}). In the soft sector we can assume $v_2^{soft}\sim c\, \varepsilon_2$ and for the hard sector since we have a $p_T$ dependent quantity one takes $v_2^{hard}\sim \chi_2(p_T) \varepsilon_2$.  Substituting them into Eq.\ (\ref{eqn:e2exp}), one finds 
\begin{equation}\label{eqn:rms}
v^{exp}_2(p_T)\sim\chi_2(p_T) \sqrt{\langle \varepsilon_2^2\rangle}.
\end{equation}
Since $\langle\varepsilon_2\rangle<\sqrt{\langle \varepsilon_2^2\rangle}$ one can see that event-by-event fluctuations will always increase the flow harmonics (soft $v_2$ is also proportional to the root mean squared not the mean \cite{Ollitrault:2009ie}).   Note that the use of Eq. (\ref{eqn:e2exp}) is not the only source of enhancement for $v_2$, rather it is the combination of full event-by-event hydrodynamically expanding backgrounds with the correct calculation of  Eq.\ (\ref{eqn:e2exp}) that leads to allows for a simultaneous description and solution of the $R_{AA}\otimes v_2$ puzzle.

\section{Results}

\begin{figure}[h]\centering
\includegraphics[width=26pc]{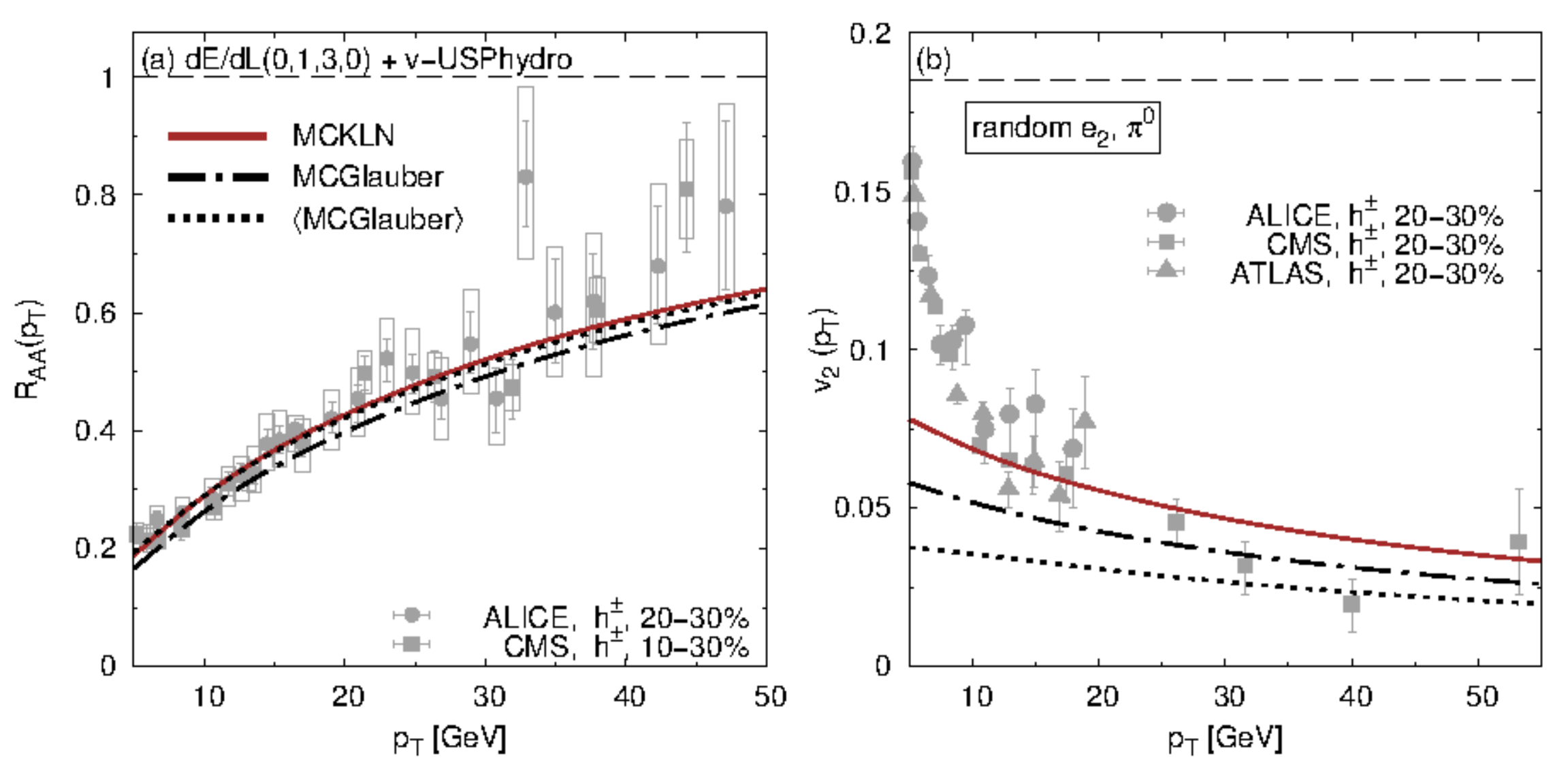}\hspace{2pc}%
\caption{\label{RAAv2}Comparison of event-by-event calculations in 
mid-central $\sqrt{s}=2.76$ TeV Pb+Pb collisions of $R_{AA}$ and $v_2$ from v-USPhydro+BBMG \cite{Noronha-Hostler:2016eow} to experimental data at LHC from ALICE  \cite{ALICE_RAA,ALICE_v2_v3}, CMS \cite{CMS_RAA,CMS_v2}, and ATLAS \cite{ATLAS_v2}.}
\end{figure}

In Fig. \ref{RAAv2} the result of the v-USPhydro+BBMG calculations are shown for $R_{AA}$ and $v_2$ compared to ALICE  \cite{ALICE_RAA,ALICE_v2_v3}, CMS \cite{CMS_RAA,CMS_v2}, and ATLAS \cite{ATLAS_v2} data  at the LHC. It is clear that the inclusion of event-by-event fluctuations with soft-hard $v_2$ correlation was a needed ingredient in order to bring up the elliptical flow to the experimental data.  In fact, one can see in Fig. \ref{RAAv2} that while $R_{AA}$ is relatively insensitive to the effect of initial conditions and event-by-event fluctuations, $v_2$ has a clear splitting between MCGlauber and MCKLN initial conditions exactly as in the soft sector (see that in Fig. \ref{softv2} the $v_2$ from MCKLN is always larger than from MCGlauber due to the larger eccentricities of MCKLN).  

\begin{figure}[h]\centering
\includegraphics[width=13pc]{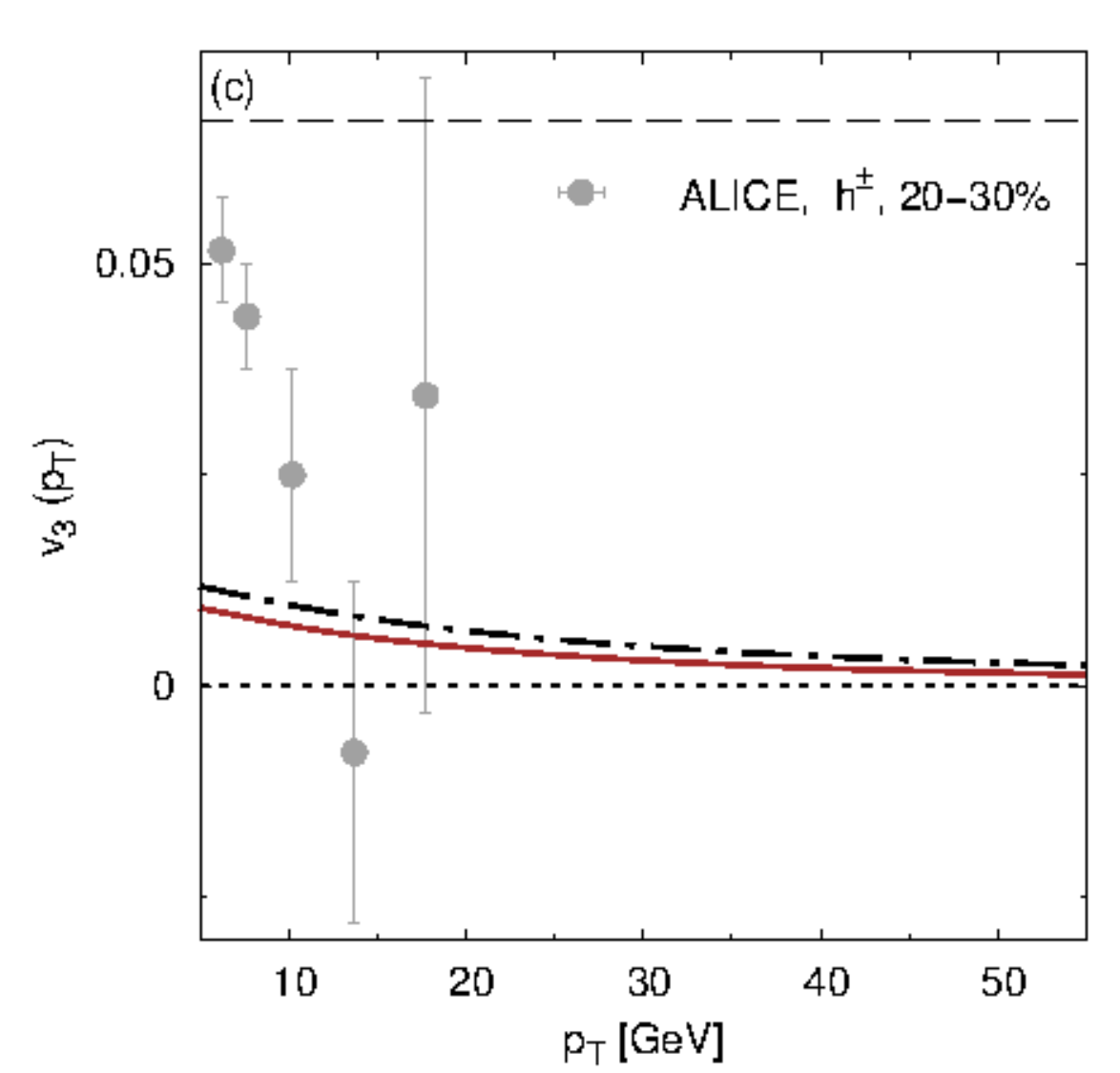}\hspace{2pc}%
\caption{\label{v3}Comparison of event-by-event calculations in 
mid-central $\sqrt{s}=2.76$ TeV Pb+Pb collisions of  $v_3$ from v-USPhydro+BBMG \cite{Noronha-Hostler:2016eow} to experimental data at LHC from ALICE  \cite{ALICE_v2_v3}.}
\end{figure}

In fact, the similarities in the soft sector also hold true for $v_3$ at high $p_T$ as seen in Fig. \ref{v3}.  MCKLN produces smaller $\varepsilon_3$ than MCGlauber, which correlates to a smaller $v_3$ both in the soft sector Fig. \ref{softv2} and in the hard sector Fig. \ref{v3}.  Furthermore, the simple existence of a non-zero $v_3$ at high $p_T$ demonstrates that event-by-event fluctuations are needed.  Finally, an interesting avenue to explore for the future is the decorrelation between the soft and hard the event-planes for triangular flow (and above) \cite{Jia:2012ez}.  

\section{Conclusions and Outlook}

In this talk, it was shown that that event-by-event fluctuations combined with the proper calculation of $v_2^{exp}$ at high $p_T$ provided a natural solution to the $R_{AA}\otimes v_2$ puzzle. Event-by-event fluctuations have been clearly measured already at both RHIC and LHC and they have been shown to be necessary to describe $v_3$ \cite{Alver:2010gr}, $v_n$ distributions \cite{Aad:2013xma}, event shape engineering \cite{Adam:2015eta}, $SC(n,m)$ correlations \cite{ALICE:2016kpq,Giacalone:2016afq}, and higher order $v_n\{m\}$ cumulants  \cite{Chatrchyan:2012ta,Abelev:2014mda,Aad:2014vba}. Thus, a proper comparison to high $p_T$ $v_2$ experimental data necessarily requires the inclusion of event-by-event hydrodynamic fluctuations as well as the calculation of the soft-hard correlation in Eq.\ (\ref{softhard}).

\begin{figure}[h]\centering
\includegraphics[width=26pc]{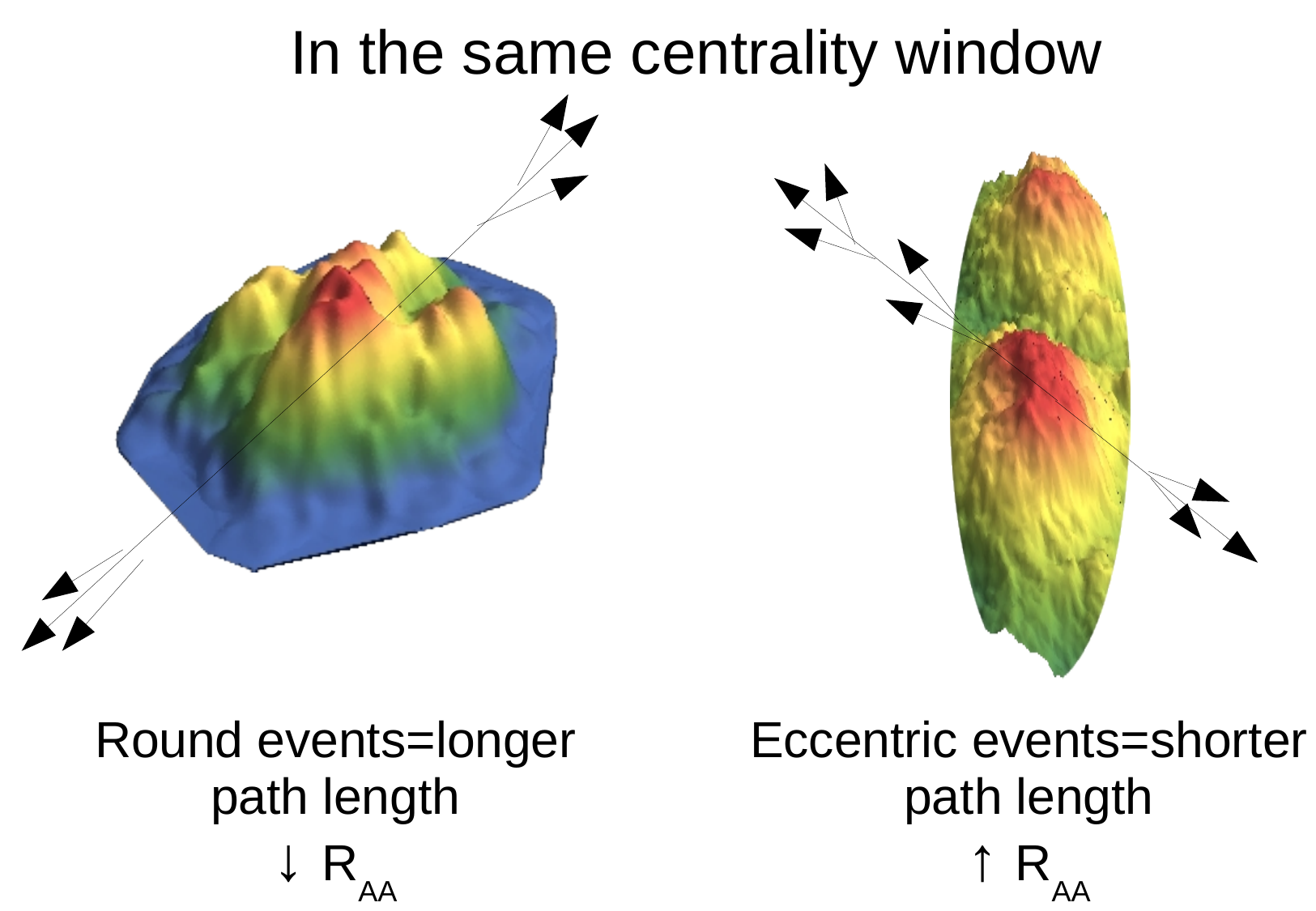}\hspace{2pc}%
\caption{\label{ese}Within the same centrality class the path length can vary strongly on an event-by-event basis producing different $R_{AA}$'s.}
\end{figure}

Once one accepts that event-by-event fluctuations also play a key role in the hard sector, it opens up an entire new field of possible calculations.  In \cite{Noronha-Hostler:2016eow} it was suggested that event-shape engineering could be preformed also in the hard sector.  Fig.\ \ref{ese} illustrates the underlying concept.  For the same centrality class, e.g. density, one has a wide range of eccentricities $\varepsilon_n$'s.  It is already known that the shape of the event determines the path length seen by the jet, which is why centrality class selection is so important.  However, using centrality class alone as an estimate for the path length is inherently flawed because within a single centrality class there are multiple possible path lengths depending on the eccentricities of the event.  Thus, one possibility is to select on events with the same soft $p_T$ flow harmonic in order to hold the path length constant within the centrality class.  Additionally, comparisons can be made while holding the density fixed and changing only the path length by selecting on a variety of soft $p_T$ flow harmonics to study the effect of energy loss for a very specific path length.  Qualitatively, one can see from Fig.\ \ref{ese} that highly eccentric events have a shorter path length (since statistically its much more likely for a jet to cross the thin part of the event vs. the long part) whereas circular events have the longest average path length.  

Experimental studies at ATLAS \cite{Aad:2015lwa} and more recently at CMS \cite{CMS:2016uwf} have already begun comparing soft-hard flow harmonics.  Both demonstrate a very clear linear correlation between the soft and hard $v_2$ (as shown in \cite{Noronha-Hostler:2016eow}) further solidifying the connection between $v_2$ and  $\sqrt{\langle \varepsilon_2^2\rangle}$ in Eq.\ (\ref{eqn:rms}).  Additionally, CMS \cite{CMS:2016uwf} found a convergence of the cumulants of flow harmonics that is a good indication that there is collective flow even at high $p_T$.  

The success of v-USPhydro+BBMG in fitting  $v_2(p_T>10 GeV)$ makes one think whether the hard flow harmonics are also just as sensitive to the usual hydrodynamic parameters such as the initial time $\tau_0$, the switching temperature $T_{sw}$, the transport coefficients etc.    While the heavy flavor shows a strong dependence on the heavy quark drag coefficient \cite{Das:2015ana}, it also shows a very strong dependence on a temperature dependent shear viscosity to entropy density ratio $\eta/s(T)$  \cite{Esha:2016svw}.  It is natural to wonder how a temperature dependent $\eta/s(T)$ would affect high $p_T$ flow given its connection with the jet transport coefficient $\hat{q}$ \cite{Majumder:2007zh}. Furthermore, using jet physics one can even extract a minimum at $T_c$ \cite{Xu:2014tda} analogous to what is found in the hadronic phase \cite{NoronhaHostler:2008ju,NoronhaHostler:2012ug}.  Since there is still a wide variation in $\eta/s(T)$ calculations \cite{Noronha-Hostler:2015qmd} any additional constraint on the values of transport coefficients is sorely needed. 

The most obvious step towards the future of combining soft and hard physics is to include hard scattering effects within relativistic hydrodynamics in order to study the intermediate $p_T$ range of flow harmonics that are not possible with only relativistic hydrodynamics or energy loss models on their own.  Initial work has been done in this direction \cite{Crkovska:2016flo,Andrade:2014swa,Pang:2012he} with the hope that this will be explored in more depth in the near future.


\section*{References}
\begin{thebibliography}{9}
\bibitem{Adare:2010sp} 
  A.~Adare {\it et al.} [PHENIX Collaboration],
  Phys.\ Rev.\ Lett.\  {\bf 105}, 142301 (2010)
  doi:10.1103/PhysRevLett.105.142301
  [arXiv:1006.3740 [nucl-ex]].

\bibitem{Wang:2000fq} 
  X.~N.~Wang,
  Phys.\ Rev.\ C {\bf 63}, 054902 (2001)
  doi:10.1103/PhysRevC.63.054902
  [nucl-th/0009019].
  
\bibitem{Gyulassy:2000gk} 
  M.~Gyulassy, I.~Vitev and X.~N.~Wang,
  Phys.\ Rev.\ Lett.\  {\bf 86}, 2537 (2001)
  doi:10.1103/PhysRevLett.86.2537
  [nucl-th/0012092].
  
\bibitem{Shuryak:2001me} 
  E.~V.~Shuryak,
  Phys.\ Rev.\ C {\bf 66}, 027902 (2002)
  doi:10.1103/PhysRevC.66.027902
  [nucl-th/0112042].
  
\bibitem{Betz:2014cza} 
  B.~Betz and M.~Gyulassy,
  JHEP {\bf 1408}, 090 (2014)
  [JHEP {\bf 1410}, 043 (2014)]
  doi:10.1007/JHEP10(2014)043, 10.1007/JHEP08(2014)090
  [arXiv:1404.6378 [hep-ph]].
  
\bibitem{Xu:2014tda} 
  J.~Xu, J.~Liao and M.~Gyulassy,
  Chin.\ Phys.\ Lett.\  {\bf 32}, no. 9, 092501 (2015)
  doi:10.1088/0256-307X/32/9/092501
  [arXiv:1411.3673 [hep-ph]].
  
\bibitem{Alver:2010gr} 
  B.~Alver and G.~Roland,
  Phys.\ Rev.\ C {\bf 81}, 054905 (2010)
  [Phys.\ Rev.\ C {\bf 82}, 039903 (2010)]
  doi:10.1103/PhysRevC.82.039903, 10.1103/PhysRevC.81.054905
  [arXiv:1003.0194 [nucl-th]].

\bibitem{Paiva:1996nv} 
  S.~Paiva, Y.~Hama and T.~Kodama,
  Phys.\ Rev.\ C {\bf 55}, 1455 (1997).
  doi:10.1103/PhysRevC.55.1455
  
\bibitem{Gyulassy:1996br} 
  M.~Gyulassy, D.~H.~Rischke and B.~Zhang,
  Nucl.\ Phys.\ A {\bf 613}, 397 (1997)
  doi:10.1016/S0375-9474(96)00416-2
  [nucl-th/9609030].

\bibitem{Aguiar:2001ac} 
  C.~E.~Aguiar, Y.~Hama, T.~Kodama and T.~Osada,
  Nucl.\ Phys.\ A {\bf 698}, 639 (2002)
  doi:10.1016/S0375-9474(01)01447-6
  [hep-ph/0106266].
  
\bibitem{Socolowski:2004hw} 
  O.~Socolowski, Jr., F.~Grassi, Y.~Hama and T.~Kodama,
  Phys.\ Rev.\ Lett.\  {\bf 93}, 182301 (2004)
  doi:10.1103/PhysRevLett.93.182301
  [hep-ph/0405181].
  
\bibitem{Andrade:2006yh} 
  R.~Andrade, F.~Grassi, Y.~Hama, T.~Kodama and O.~Socolowski, Jr.,
  Phys.\ Rev.\ Lett.\  {\bf 97}, 202302 (2006)
  doi:10.1103/PhysRevLett.97.202302
  [nucl-th/0608067].
  
  


  
  
\bibitem{Gardim:2011xv} 
  F.~G.~Gardim, F.~Grassi, M.~Luzum and J.~Y.~Ollitrault,
  Phys.\ Rev.\ C {\bf 85}, 024908 (2012)
  [arXiv:1111.6538 [nucl-th]].
  
\bibitem{Gardim:2014tya} 
  F.~G.~Gardim, J.~Noronha-Hostler, M.~Luzum and F.~Grassi,
  Phys.\ Rev.\ C {\bf 91}, no. 3, 034902 (2015)
  doi:10.1103/PhysRevC.91.034902
  [arXiv:1411.2574 [nucl-th]].

      
\bibitem{Drescher:2006ca} 
  H.-J.~Drescher and Y.~Nara,
  Phys.\ Rev.\ C {\bf 75}, 034905 (2007)
  doi:10.1103/PhysRevC.75.034905
  [nucl-th/0611017].
  
\bibitem{Noronha-Hostler:2015coa} 
  J.~Noronha-Hostler, J.~Noronha and M.~Gyulassy,
  Phys.\ Rev.\ C {\bf 93}, no. 2, 024909 (2016)
  doi:10.1103/PhysRevC.93.024909
  [arXiv:1508.02455 [nucl-th]].
  
\bibitem{Aad:2013xma} 
  G.~Aad {\it et al.} [ATLAS Collaboration],
  JHEP {\bf 1311}, 183 (2013)
  doi:10.1007/JHEP11(2013)183
  [arXiv:1305.2942 [hep-ex]].
  
  
\bibitem{Noronha-Hostler:2015uye} 
  J.~Noronha-Hostler, M.~Luzum and J.~Y.~Ollitrault,
  Phys.\ Rev.\ C {\bf 93}, no. 3, 034912 (2016)
  doi:10.1103/PhysRevC.93.034912
  [arXiv:1511.06289 [nucl-th]].
  
  
\bibitem{Betz:2015mlf} 
  B.~Betz, F.~Senzel, C.~Greiner and M.~Gyulassy,
  arXiv:1512.07443 [hep-ph].
  
\bibitem{Crkovska:2016flo} 
  J.~Crkovska {\it et al.},
  arXiv:1603.09621 [hep-ph].
  
\bibitem{Zhang:2012ha} 
  X.~Zhang and J.~Liao,
  Phys.\ Rev.\ C {\bf 87}, 044910 (2013)
  doi:10.1103/PhysRevC.87.044910
  [arXiv:1210.1245 [nucl-th]].
  
  
  
  
  
\bibitem{Noronha-Hostler:2016eow} 
  J.~Noronha-Hostler, B.~Betz, J.~Noronha and M.~Gyulassy,
  arXiv:1602.03788 [nucl-th].
  
 
  
\bibitem{Noronha-Hostler:2013gga} 
  J.~Noronha-Hostler, G.~S.~Denicol, J.~Noronha, R.~P.~G.~Andrade and F.~Grassi,
  Phys.\ Rev.\ C {\bf 88}, 044916 (2013)
  [arXiv:1305.1981 [nucl-th]].
  
  
\bibitem{Noronha-Hostler:2014dqa} 
  J.~Noronha-Hostler, J.~Noronha and F.~Grassi,
  Phys.\ Rev.\ C {\bf 90}, no. 3, 034907 (2014)
  [arXiv:1406.3333 [nucl-th]].
  
\bibitem{Betz:2011tu} 
  B.~Betz, M.~Gyulassy and G.~Torrieri,
  Phys.\ Rev.\ C {\bf 84}, 024913 (2011)
  doi:10.1103/PhysRevC.84.024913
  [arXiv:1102.5416 [nucl-th]].
  
\bibitem{Betz:2012qq} 
  B.~Betz and M.~Gyulassy,
  Phys.\ Rev.\ C {\bf 86}, 024903 (2012)
  doi:10.1103/PhysRevC.86.024903
  [arXiv:1201.0281 [nucl-th]].
  
\bibitem{Luzum:2013yya} 
  M.~Luzum and H.~Petersen,
  J.\ Phys.\ G {\bf 41}, 063102 (2014)
  doi:10.1088/0954-3899/41/6/063102
  [arXiv:1312.5503 [nucl-th]].
  

  
\bibitem{Noronha-Hostler:2015dbi} 
  J.~Noronha-Hostler, L.~Yan, F.~G.~Gardim and J.~Y.~Ollitrault,
  Phys.\ Rev.\ C {\bf 93}, no. 1, 014909 (2016)
  doi:10.1103/PhysRevC.93.014909
  [arXiv:1511.03896 [nucl-th]].
  
  
\bibitem{Ollitrault:2009ie} 
  J.~Y.~Ollitrault, A.~M.~Poskanzer and S.~A.~Voloshin,
  Phys.\ Rev.\ C {\bf 80}, 014904 (2009)
  doi:10.1103/PhysRevC.80.014904
  [arXiv:0904.2315 [nucl-ex]].
  
  \bibitem{ALICE_RAA}
B.~Abelev {\it et al.}  [ALICE Collaboration], Phys.\ Lett.\ B {\bf
720}, 52 (2013).

\bibitem{CMS_RAA}
S.~Chatrchyan {\it et al.}  [CMS Collaboration], Eur.\ Phys.\ J.\ C {\bf
72}, 1945 (2012).

\bibitem{ALICE_v2_v3}
B.~Abelev {\it et al.}  [ALICE Collaboration], Phys.\ Lett.\ B {\bf
719}, 18 (2013).

\bibitem{CMS_v2}
S.~Chatrchyan {\it et al.}  [CMS Collaboration], Phys.\ Rev.\ Lett.\
{\bf 109}, 022301 (2012).

\bibitem{ATLAS_v2}
G.~Aad {\it et al.}  [ATLAS Collaboration], Phys.\ Lett.\ B {\bf 707},
330 (2012).  

\bibitem{Jia:2012ez} 
  J.~Jia,
  Phys.\ Rev.\ C {\bf 87}, no. 6, 061901 (2013)
  doi:10.1103/PhysRevC.87.061901
  [arXiv:1203.3265 [nucl-th]].
  
\bibitem{Adam:2015eta} 
  J.~Adam {\it et al.} [ALICE Collaboration],
  Phys.\ Rev.\ C {\bf 93}, no. 3, 034916 (2016).
  
\bibitem{ALICE:2016kpq} 
  J.~Adam {\it et al.} [ALICE Collaboration],
  arXiv:1604.07663 [nucl-ex].
\bibitem{Giacalone:2016afq} 
  G.~Giacalone, L.~Yan, J.~Noronha-Hostler and J.~Y.~Ollitrault,
  arXiv:1605.08303 [nucl-th].
  

\bibitem{Chatrchyan:2012ta} 
  S.~Chatrchyan {\it et al.} [CMS Collaboration],
  Phys.\ Rev.\ C {\bf 87}, no. 1, 014902 (2013)
  doi:10.1103/PhysRevC.87.014902
  [arXiv:1204.1409 [nucl-ex]].
\bibitem{Abelev:2014mda} 
  B.~B.~Abelev {\it et al.} [ALICE Collaboration],
  Phys.\ Rev.\ C {\bf 90}, no. 5, 054901 (2014)
  doi:10.1103/PhysRevC.90.054901
  [arXiv:1406.2474 [nucl-ex]].
\bibitem{Aad:2014vba} 
  G.~Aad {\it et al.} [ATLAS Collaboration],
  Eur.\ Phys.\ J.\ C {\bf 74}, no. 11, 3157 (2014)
  doi:10.1140/epjc/s10052-014-3157-z
  [arXiv:1408.4342 [hep-ex]].
  
\bibitem{Aad:2015lwa} 
  G.~Aad {\it et al.} [ATLAS Collaboration],
  Phys.\ Rev.\ C {\bf 92}, no. 3, 034903 (2015)
  doi:10.1103/PhysRevC.92.034903
  [arXiv:1504.01289 [hep-ex]].
  
\bibitem{CMS:2016uwf} 
  CMS Collaboration [CMS Collaboration],
  CMS-PAS-HIN-15-014.
  
\bibitem{Das:2015ana} 
  S.~K.~Das, F.~Scardina, S.~Plumari and V.~Greco,
  Phys.\ Lett.\ B {\bf 747}, 260 (2015)
  doi:10.1016/j.physletb.2015.06.003
  [arXiv:1502.03757 [nucl-th]].
  
\bibitem{Esha:2016svw} 
  R.~Esha, M.~Nasim and H.~Z.~Huang,
  arXiv:1603.02700 [nucl-th].
  
  
\bibitem{Majumder:2007zh} 
  A.~Majumder, B.~Muller and X.~N.~Wang,
  Phys.\ Rev.\ Lett.\  {\bf 99}, 192301 (2007)
  doi:10.1103/PhysRevLett.99.192301
  [hep-ph/0703082].
  

  
\bibitem{NoronhaHostler:2008ju} 
  J.~Noronha-Hostler, J.~Noronha and C.~Greiner,
  Phys.\ Rev.\ Lett.\  {\bf 103}, 172302 (2009)
  doi:10.1103/PhysRevLett.103.172302
  [arXiv:0811.1571 [nucl-th]].
\bibitem{NoronhaHostler:2012ug} 
  J.~Noronha-Hostler, J.~Noronha and C.~Greiner,
  Phys.\ Rev.\ C {\bf 86}, 024913 (2012)
  doi:10.1103/PhysRevC.86.024913
  [arXiv:1206.5138 [nucl-th]].
  
\bibitem{Noronha-Hostler:2015qmd} 
  J.~Noronha-Hostler,
  arXiv:1512.06315 [nucl-th].
  
\bibitem{Andrade:2014swa} 
  R.~P.~G.~Andrade, J.~Noronha and G.~S.~Denicol,
  Phys.\ Rev.\ C {\bf 90}, no. 2, 024914 (2014)
  doi:10.1103/PhysRevC.90.024914
  [arXiv:1403.1789 [nucl-th]].
  
\bibitem{Pang:2012he} 
  L.~Pang, Q.~Wang and X.~N.~Wang,
  Phys.\ Rev.\ C {\bf 86}, 024911 (2012)
  doi:10.1103/PhysRevC.86.024911
  [arXiv:1205.5019 [nucl-th]].

  
\end{thebibliography}
\end{document}